\documentclass[pre,twocolumn,floatfix,amsmath,amsfonts,longbibliography]{revtex4}
\usepackage{graphicx,epstopdf,url}
\usepackage{color}
\usepackage[colorlinks=true,urlcolor=blue,citecolor=blue,linkcolor=blue,urlcolor=blue]{hyperref}
\usepackage[active]{srcltx}
\begin{document}

\title{
Shock absorption by multilayer carbon nanotube packings
}

\author{Alexander V. Savin}
\email{asavin@chph.ras.ru}
\affiliation{
N.N. Semenov Federal Research Center for Chemical Physics of the Russian Academy of Sciences, 4 Kosygin Street, Moscow 119991, Russia}
\affiliation{
Plekhanov Russian University of Economics, 36 Stremyanny Lane, Moscow 117997, Russia
}

\begin{abstract}
The propagation of transverse impact energy in a multilayer packing (in an array)
of parallel single-walled carbon nanotubes has been simulated.
It has been shown that such nanotube arrays are effective shock absorbers.
The depreciation effect is most pronounced for packings of nanotubes with a diameter of 2.7--3.9~nm.
Here, part of the impact energy is absorbed due to the transfer of the packing to a higher energy stationary state, in which part of the nanotubes is in a collapsed state.
The impact impulse reaches the other edge of the packing most weakened and distributed over time.
For nanotubes with a smaller diameter, the compression of the array occurs elastically without energy accumulation, and for nanotubes with a larger diameter -- with energy release.

\end{abstract}
\maketitle

\section{Introduction \label{sec1}}
Carbon nanotubes (CNTs) have the shape of a graphene sheet rolled into a cylinder with a diameter greater than 0.4 nanometers and a length of up to tens of centimeters.
For the first time, topologically similar molecular structures were obtained by thermal decomposition of carbon monoxide on an iron contact \cite{Radush52}.
CNTs themselves, as ideal cylindrical macromolecules, were obtained much later as byproducts of the synthesis of fullerene C$_{60}$ \cite{Iijima91a,Iijima91b}.
CNTs are currently attracting increased attention due to their unique mechanical properties \cite{Treacy1996,Wong1997}, high thermal conductivity \cite{Kim2001,Pop2005,Fujii2005,Chang2007} and electrical conductivity \cite{Popov2004,Hecht2011}.
Currently, nanotubes with the required geometric properties (i.e., with the required diameter, length, and chirality) can be easily synthesized \cite{Di2016, Bai2018} and used to produce parallel CNT bundles \cite{Liu2003,Li2005}.
Such materials, also called forests or CNT arrays, have even more interesting mechanical properties compared to isolated nanotubes due to the van der Waals interactions between them \cite{Rakov2013}.

Nanotubes have a high longitudinal (axial) and relatively weak transverse (radial) rigidity.
Because of this, with a sufficiently large diameter, a nanotube can transition from a hollow cylindrical shape to a collapsed state due to the non-valent interaction of atoms of its surface \cite{Chopra95,Gao98,Xiao07,Chang2008,Baimova15,Impellizzeri2019,Maslov2020}.
The critical diameter value for a single CNT may decrease for nanotubes in the bundle \cite{Drozdov2019,Onuoha2020}.
Non-valent interaction with the substrate can also lead to a change in the cylindrical shape of the nanotube \cite{Hertel98,Xie10,Yuan18}.

Modeling of the transverse compression of multilayer packing of identical parallel single-walled CNTs shows that with a diameter of nanotubes $D>2.5$~nm, their array is a multistable system with many stable stationary states, differing from each other in the proportions of nanotubes in the collapsed state \cite{Savin2022pss,Savin2022jetp}.
Therefore, the energy of the nanotube bundle, depending on the value of their diameter, can not only increase, but also decrease during transverse compression.
It can be said that nanotube bundles (multilayer packings) are nanoscale analogues of mechanical metamaterials with negative stiffness \cite{Darwish2024,Tan2025}.
Metamaterials (cellular structures) showing negative stiffness can be used as highly efficient energy dissipation systems.

The high compressibility of CNT arrays allows them to be used to protect against shocks and vibrations \cite{Cao2005,Rysaeva2020}.
Composite materials reinforced with carbon nanotubes very effectively absorb the energy of shock waves and vibrations \cite{Abdullah2019,Moumen2019,Park2020,Kim2021,Vinyas2021}, and also protect against heat shock \cite{Liu2020,Pourasghar2022}.

In this paper, we will simulate the scattering of transversal-impact energy by a multilayer nanotube packing with a chirality index $(m,0)$.
Using a two-dimensional model of the cyclic molecular chain system \cite{Savin2015prb,Savin2017cms}, it will be shown that a system of multilayer CNT packings can serve as a highly effective shock-absorbing layer.
Packings (arrays) of single-walled nanotubes with a diameter of $2.7<D<3.9$~nm provide the best cushioning.

This paper is organized as follows:
In Sec. \ref{sec2} a 2D chain model is constructed, which is then used to model deformations of multilayer nanotube packings.
Section \ref{sec3} is devoted to the behavior of a packing when it is laterally compressed.
Sec. \ref{sec4} describes the possible stationary states of packings.
The absorption of the energy of an external impact by a multilayer packing is modeled in sec. \ref{sec5}.
The conclusions and conclusions are given in sec. \ref{sec6}.
\begin{figure}[tb]
\begin{center}
\includegraphics[angle=0, width=1.0\linewidth]{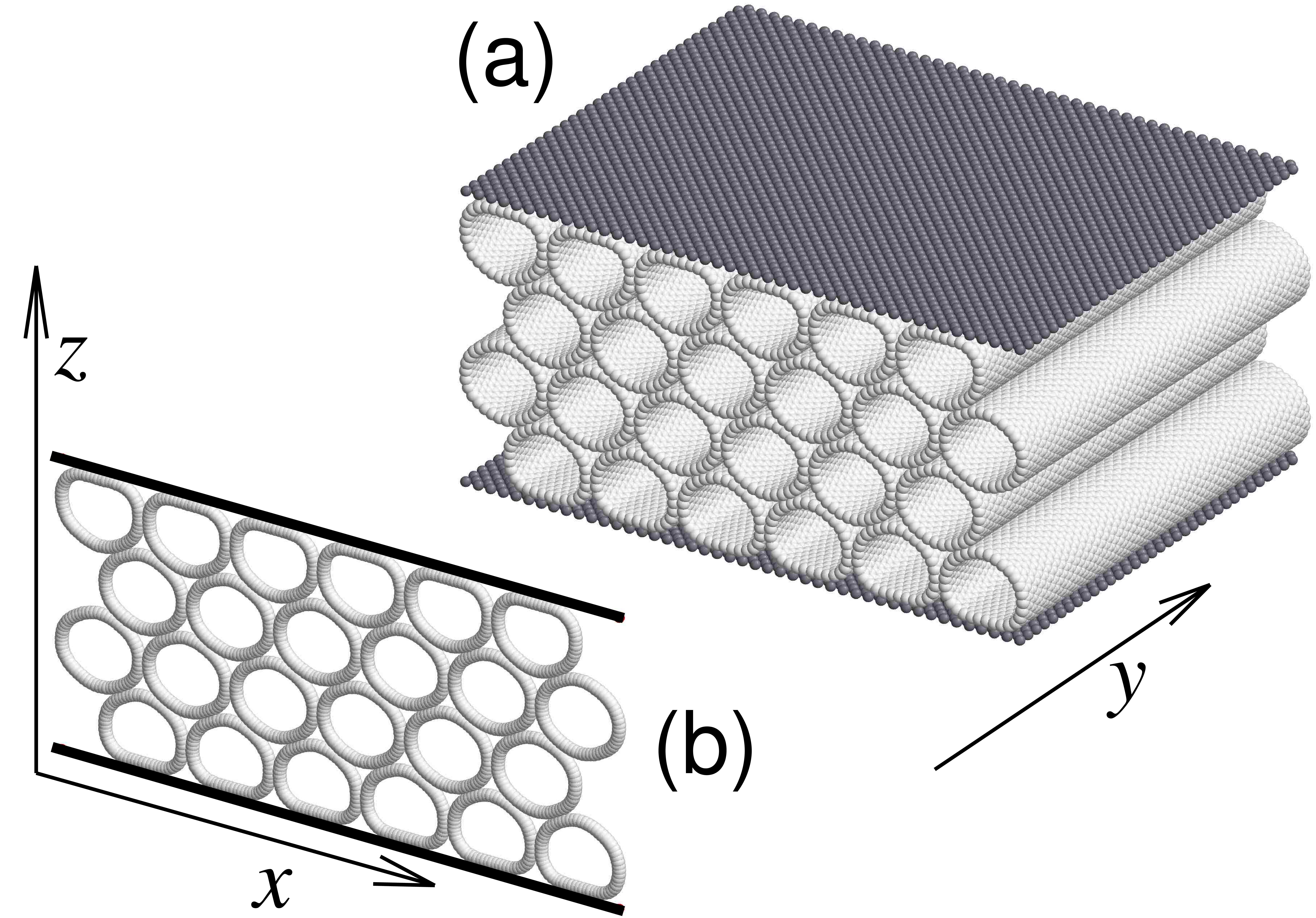}
\end{center}
\caption{\label{fg01}\protect
Scheme for constructing a two-dimensional model of a system of parallel CNTs lying between two flat substrates: (a) a full-atomic model of a packing of $6\times 4$ nanotubes with a chirality index (30,0) lying between flat surfaces of silver crystals and (b) a cross-section of the system (thick straight lines show the position of the surface of flat substrates).
Compression of the CNT system (convergence of substrates) is carried out along the $z$ axis, periodic boundary conditions are used along the $x$ axis.
For the presented system, the number of nanotubes in one layer is $N_x=6$, the number of layers is $N_z=4$, and the period along the $x$ axis is $a_x=16.98$~nm.
}
\end{figure}

\section{2D model of a carbon nanotube array \label{sec2}}
Consider a system of parallel $N_{xz}=N_xN_z$ single-walled CNTs with chirality index $(m,0)$
located between two flat substrates ($N_x$ is the number of nanotubes in one layer parallel to the substrate, $N_z$ is the number of layers) -- see Fig. \ref{fg01} (a).
We will assume that all nanotubes are located parallel to the $y$ axis, and the planes of the bounding substrates are parallel to the $(x,y)$ plane.
Along the $x$ axis, we will use periodic boundary conditions with a period $a_x$ (the period value depends on the values of the packing parameters $m$, $N_x$, $N_z$).
The homogeneous compression of a system of parallel nanotubes along the $z$ axis (transverse compression) is conveniently described using a two-dimensional model of a cyclic molecular chain system \cite{Savin2015prb,Savin2017cms}.

For a single-walled CNT with a zigzag structure [with a chirality index $(m,0)$], the chain model describes a nanotube cross-section forming an annular chain of $N=2m$ effective particles (united atoms) corresponding to the longitudinal lines of atoms in the nanotube -- see Fig. \ref{fg01} (b).

Under uniform compression, the shape of the nanotube's cross-section fully describes its deformation.
Therefore, compression of a system of parallel nanotubes can be described as deformation of their cross-sections.
This approach makes it possible to significantly reduce the dimension of the simulated molecular system.
This model has previously been used to simulate the transverse compression of a nanotube bundle \cite{Korznikova2019}, to analyze the mechanical properties of single-walled and multi-walled nanotubes located on flat substrates \cite{Savin2019ftt,Savin2021ftt,Savin2022pss,Savin2022jetp} and for modeling the transverse thermal conductivity of a nanotube array \cite{Savin2022ftt}.

Let us consider a bundle of straight, single-walled CNTs$(m,0)$ oriented along the $y$ axis and located between two parallel flat surfaces $z=h_1$ and $z=h_2$ ($h_2>h_1$) -- see Fig. \ref{fg01}.
The cross-section of each nanotube consists of $N=2m$ particles, each of which describes the displacements of a straight line of carbon atoms normal to the $(x,z)$-plane.
The Hamiltonian of the CNT cross-section (a cyclic chain of $N$ atoms) has the form
\begin{eqnarray}
H=\sum_{n=1}^{N}[\frac12M(\dot{\bf u}_n,\dot{\bf u}_n)+V(R_n)+U(\theta_n)
\nonumber\\
+W_0(z_n-h_1)+W_0(h_2-z_n)+\frac12\sum^N_{l=1\above 0pt |l-n|>4} W_1(r_{n,l})],
\label{f1}
\end{eqnarray}
where the 2D vector ${\bf u}_n=(x_n,z_n)$ defines the coordinates of the $n$th particle in the cycle chain, and $M=12m_p$ is the carbon atom mass ($m_p=1.6603\times 10^{-27}$~kg is proton mass).

The potential
\begin{equation}
V(R)=\frac12K(R-R_0)^2,
\label{f2}
\end{equation}
describes the longitudinal chain stiffness, where $K$ is the interaction stiffness,
$R_0$ is the equilibrium bond length (chain period), and $R_n=|{\bf v}_{n}|$
is the distance between adjacent $n$ and $n+1$ particles
(vector ${\bf v}_n={\bf u}_{n+1}-{\bf u}_n$).
\begin{table*}[tb]
\caption{
The values of the parameters of multilayer CNT$(m,0)$ packings: $N_x$ is the number of nanotubes in one layer, $N_z$ is the number of layers, $a_x$ is the period along the $x$ axis, $\Delta h_o=h_2-h_1$ is the thickness of the packing for a stationary state with all open nanotubes, $\Delta h_c$ is the thickness of the packing with the largest proportion of collapsed nanotubes.
\label{tab1}
}
\begin{center}
\begin{tabular}{cccccccccccccc}
\hline\hline
$m$              & 20    & 25    & 30     & 35     & 40     & 45     & 50     & 55     & 60     & 65&70&75&80\\
$N_x$            & 32    & 26    & 22     & 20     & 17     & 16     & 14     & 13     & 12     & 11&10&10&10\\
$N_z$            & 50    & 49    & 48     & 46     & 47     & 44     & 46     & 49     & 44     & 45&46&42&40\\
~~$a_x$ (nm)       & 60.02 & 58.88 & 58.34  & 60.52  & 57.11  & 58.82  & 55.67  & 55.72  & 55.35  & 54.19&52.39&55.70&58.95\\
$\Delta h_o$ (nm) &~81.37~&~95.94~&~109.62~&~120.14~&~139.27~&~145.69~&~168.27~&~196.05~&~190.50~&~210.14~&~230.56~&~224.18~&~226.53\\
$\Delta h_c$ (nm) &   -  &  -    & 105.38 & 71.13  & 67.91  & 63.93  &  66.15 & 69.56  & 61.91  &63.31&64.90&58.79&55.71\\
\hline\hline
\end{tabular}
\end{center}
\end{table*}

The potential
\begin{equation}
U(\theta)=\epsilon_\theta[1+\cos(\theta)],
\label{f3}
\end{equation}
describes the bending stiffness of the chain, where $\theta$ is the angle between two adjacent bonds,
cosine of the $n$th "valence"\ angle
$\cos(\theta_n)=-({\bf v}_{n-1},{\bf v}_n)/R_{n-1}R_n$.

The parameters of potentials (\ref{f2}) and (\ref{f3}) were determined in ref. \cite{Savin2015prb,Savin2015ftt} by analyzing dispersion curves for a graphene nanoribbon: longitudinal stiffness
$K=405$~N/m, chain period $R_0=r_c\sqrt{3}/2$
(where $r_c=1.418$~\AA~ is the length of C--C valence bond in the graphene sheet, $R_0=1.228$~\AA),
and energy $\epsilon_\theta=3.5$~eV.
The diameter of an isolated CNT$(m,0)$ is $D=R_0/\sin(\pi/2m)\approx 2m R_0/\pi$.

In the chain Hamiltonian (\ref{f1}) potential $W_0(h)$ describes the interaction between a chain node and the substrate formed by the flat surface of a molecular crystal ($h$ is the distance from the chain node to the substrate plane).
Let the flat substrate fill the half-space $z\le 0$.
In this case, the energy of the interaction of the carbon atom with the substrate can be described using the Lennard-Jones potential (3,9) \cite{Aitken2010,Zhang2013,Zhang2014}
\begin{equation}
Z({\bf u})=Z(z)=\varepsilon_0[(h_0/z)^9-3(h_0/z)^3]/2,
\label{f4}
\end{equation}
where $\varepsilon_0$ is the energy of bonding between a carbon atom and a substrate, $h_0$ is the equilibrium distance  from the surface plane of the substrate.
For the surface (111) of a silver crystal, energy $\varepsilon_0=0.073$~eV, distance $h_0=3$~\AA~ \cite{Tesch2016}.

Potential $W_1(r_{n,l})$ describes weak noncovalent interactions between remote nodes $n$ and $l$ of the chain, where $r_{n,l}=|{\bf u}_l-{\bf u}_n|$ is the distance between the nodes.
This potential will also be used to describe the interaction between nodes of different chains (different nanotubes).
The energy of noncovalent interaction between chain nodes can be described with high accuracy \cite{Savin2019prb}
by the (5,11) Lennard-Jones potential
\begin{equation}
W_1(r)=\epsilon_1[5(r_0/r)^{11}-11(r_0/r)^5]/6, \label{f5}
\end{equation}
with equilibrium bond length $r_0=3.607$~\AA~ and interaction energy $\epsilon_1=0.00832$~eV.
\begin{figure*}[tb]
\begin{center}
\includegraphics[angle=0, width=0.99\linewidth]{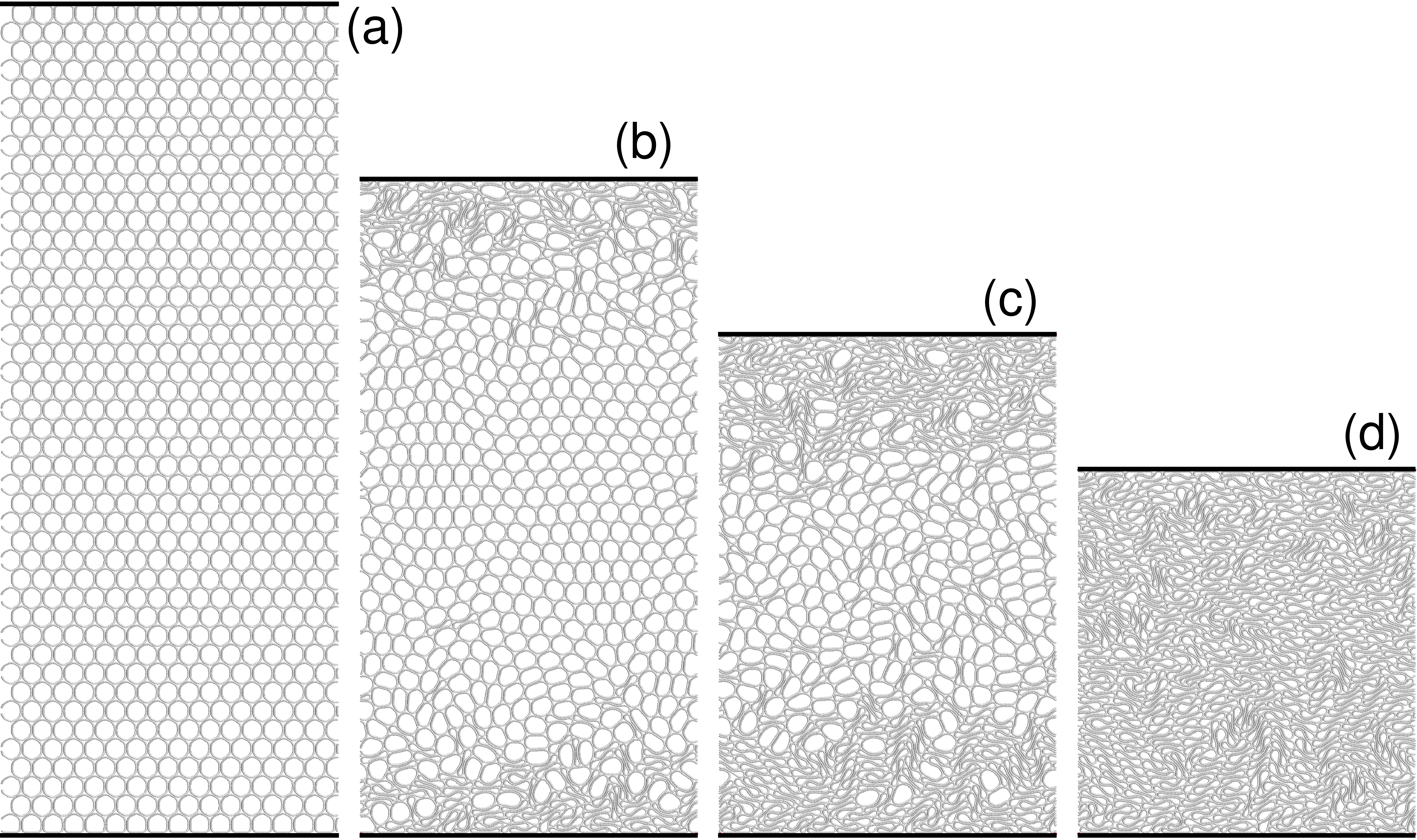}
\end{center}
\caption{\label{fg02}\protect
Steady-state packing of parallel $N_x\times N_z$ CNTs (45,0) at the fraction of collapsed nanotubes (a) $p=0$, (b) 0.34, (c) 0.68 and (d) 1.0 (packing thickness $\Delta h=145.7$, 114.9, 87.7 and 63.9~nm).
One periodic cell is shown, $N_x=16$, $N_z=44$, period $a_x=58.82$~nm.
}
\end{figure*}

Let the node coordinates of the $k$th nanotube ($k$th chain) be defined by $2N$-dimensional vector
${\bf x}_k=\{(x_{k,n},z_{k,n})\}_{n=1}^N$, $N=2m$, $k=1,...,N_{xz}$.
Then, the energy of nanotube deformation
\begin{eqnarray}
P_1({\bf x}_k)=\sum_{n=1}^N[V(R_{k,n})+U(\theta_{k,n})+W_0(z_{k,n}-h_1)\nonumber\\
+W_0(h_2-z_{k,n})+\frac12\sum^N_{l=1\above 0pt |l-n|>4} W_1(r_{k,n,l})],
\label{f6}
\end{eqnarray}
where $h_1$ and $h_2$ define the positions of the flat substrates (the distance between the substrates is $\Delta h=h_2-h_1$).

The energy of interaction of two nanotubes with coordinates
${\bf x}_{k_i}=\{ {\bf u}_{k_i,n}\}_{n=1}^N$, $i=1,2$,
\begin{equation}
P_2({\bf x}_{k_1},{\bf x}_{k_2})=\sum_{n_1=1}^N\sum_{n_2=1}^N W_1(r_{k_1,n_1;k_2,n_2}),
\label{f7}
\end{equation}
where the distance between the nodes of the chains $r_{k_1,n_1;k_2,n_2}=|{\bf u}_{k_2,n_2}-{\bf u}_{k_1,n_1}|$.

The potential energy of packing nanotubes, taking into account the periodic boundary condition along the $x$ axis, will have the form
\begin{equation}
E_p=\sum_{k=1}^{N_{xz}}P_1({\bf x}_k)+\sum_{k_1=1}^{N_{xz}-1}\sum_{k_2=k_1+1}^{N_{xz}}
[P_2({\bf x}_{k_1},{\bf x}_{k_2}+ja_x{\bf e}_x)],
\label{f8}
\end{equation}
where the vector ${\bf e}_x=\{(1,0)\}_{ n=1}^N$, and the value of $j=-1,0,1$ is chosen so that the distance between the chains with coordinates ${\bf x}_{k_1}$ and ${\bf x}_{k_2}+ja_x{\bf e}_x$ is minimal.
Note that the formula (\ref{f8}) specifies the energy of a multilayer CNT packing per its section along the $y$ axis of width $\Delta_y=3r_c/2$.

To find the stationary state of the multilayer packing of open nanotubes with free substrates (in the absence of transverse compression), it is necessary to solve the problem of minimizing the energy of the system
\begin{equation}
E_p\rightarrow\min:\{ {\bf x}_k\}_{k=1}^{N_{xz}},a_x,h_1,h_2.
\label{f9}
\end{equation}
The minimum problem (\ref{f9}) has been solved numerically by the conjugate gradient method \cite{Fletcher1964,Shanno1976}.
To obtain a stationary state with all open (uncollapsed) nanotubes, an initial configuration corresponding to the $N_z$-layered packing of $N_x$ cyclic chains in the shape of a circle has been used.
Note that the period of the calculation cell $a_x$ and the thickness of the packing $\Delta h=h_2-h_1$ depend on the values of the parameters $m$, $N_x$, $N_z$, so they have also been found when solving the problem (\ref{f9}).
A typical view of the stationary state of the layered structure of open nanotubes $\{{\bf x}_k^o\}_{k=1}^{N_{xz}}$ is shown in Fig. \ref{fg02} (a).
The values of the parameters of multilayer CNT$(m,0)$ packings  $m$, $N_x$, $N_z$ and $a_x$ used further in dynamics modeling are given in the table \ref{tab1}.
Note that it is not possible to obtain a complete match of the values of the period $a_x$ for multilayer packings of nanotubes of different diameters.
As can be seen from the table, we will use packing parameters for which the average value of the period is $a_x\approx 57.05$~nm, and the average value of the longitudinal density of atoms is $2mN_xN_z/a_x\approx 1125$~nm$^{-1}$.
\begin{figure}[tb]
\begin{center}
\includegraphics[angle=0, width=1.0\linewidth]{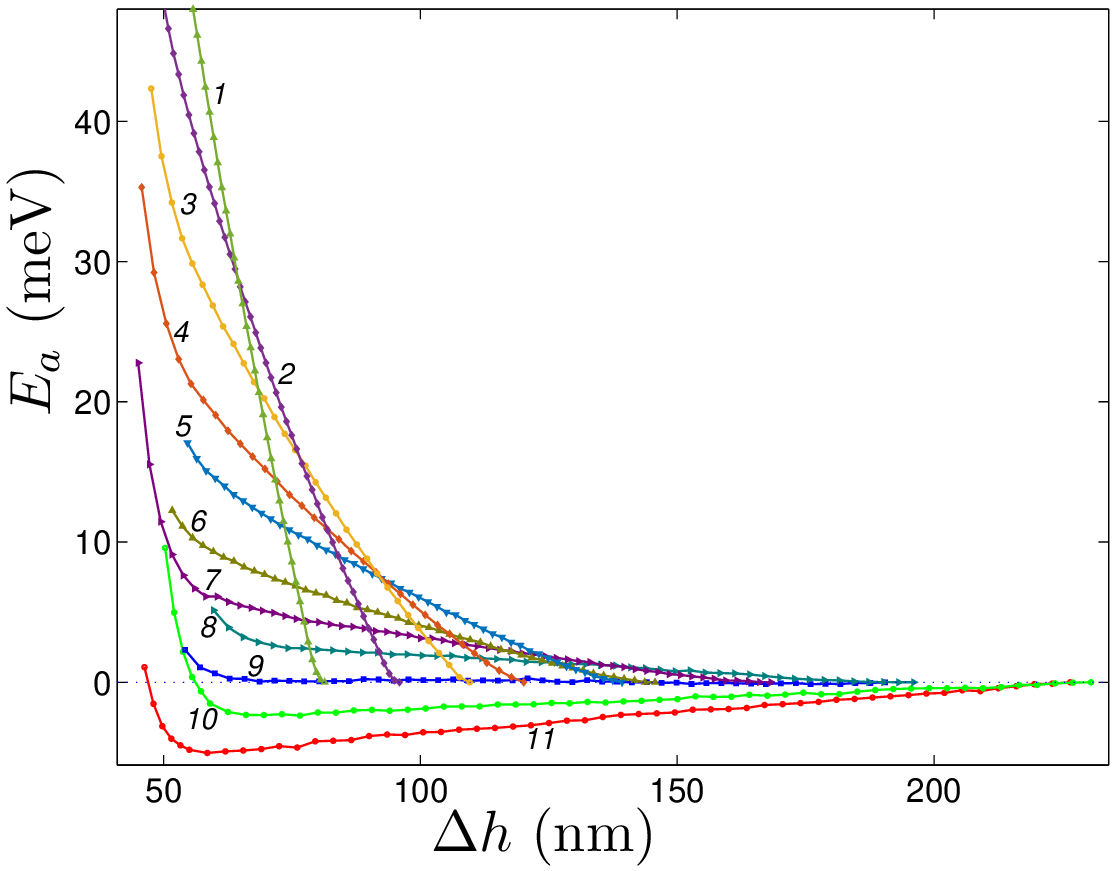}
\end{center}
\caption{\label{fg03}\protect
Dependence of the average specific energy $E_a=\bar{E}/N_a$ on the thickness $\Delta h$ of the layered $N_x\times N_z$ CNT$(m,0)$ packing with $m=20$, 25, 30,..., 60, 70, 80 (curves 1, 2, 3,...,9, 10, 11).
The energy of a state with a thickness of $\Delta h_o$ is used as the zero level.
The number of atoms in a periodic cell $N_a=2mN_xN_z$, temperature $T=300$K.
}
\end{figure}
\begin{figure}[tb]
\begin{center}
\includegraphics[angle=0, width=1.0\linewidth]{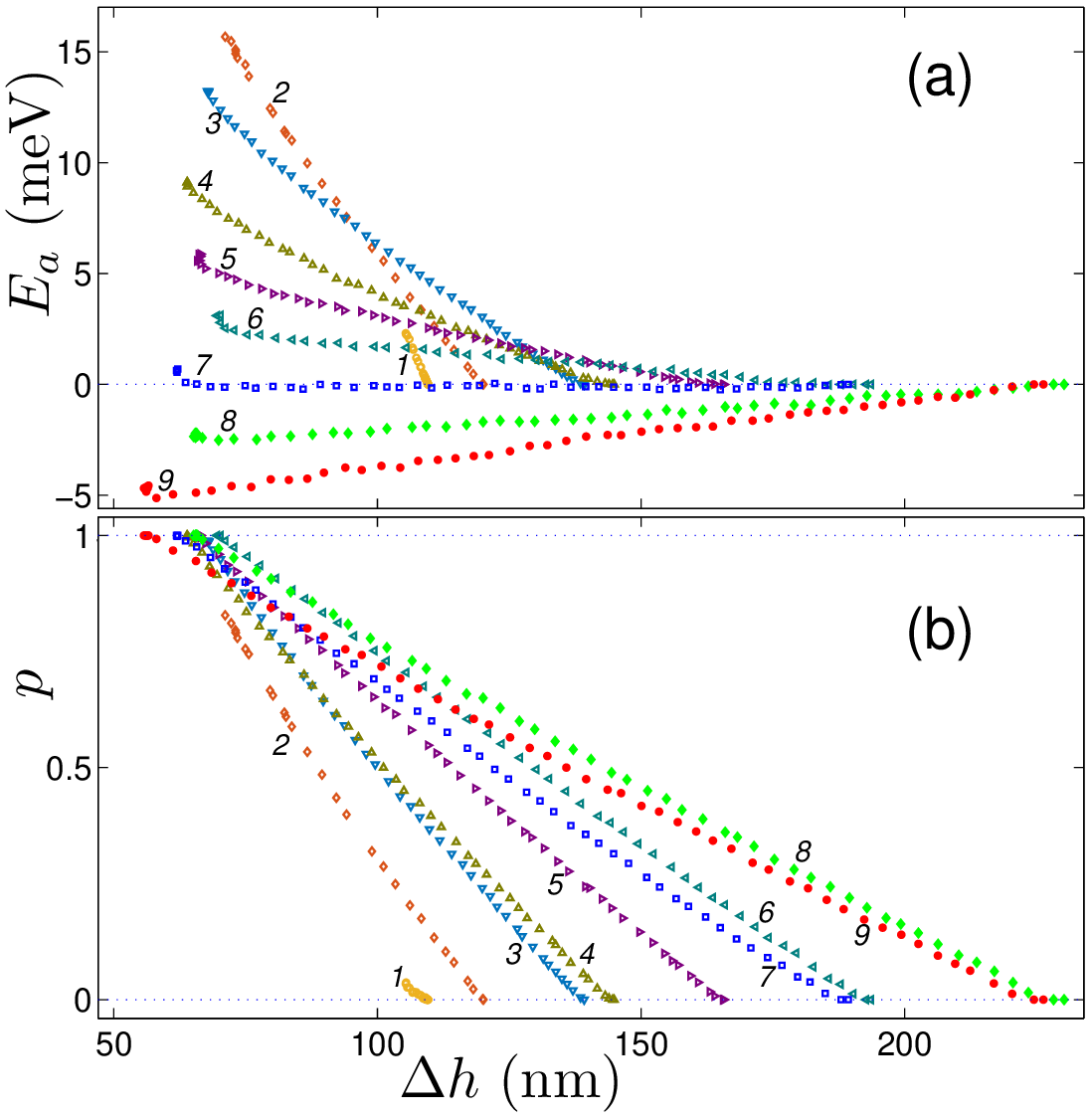}
\end{center}
\caption{\label{fg04}\protect
Dependence of (a) the specific energy $E_a=E/N_a$ and (b) the proportion of collapsed nanotubes $p$ on the thickness $\Delta h$ of the stationary state of the layered packing $N_x\times N_z$ CNT$(m,0)$ with $m=30$, 35, 40,..., 60, 70, 89 ( markers 1, 2, 3,...,7, 8, 9).
The number of atoms in a periodic cell $N_a=2mN_xN_z$.
The energy of a state with a thickness of $\Delta h_o$ at which $p=0$ is used as the zero level.
}
\end{figure}

\section{Behavior of multilayer CNT packing during its transverse compression \label{sec3}}

Let us first consider the behavior of multilayer CNT packing during its transverse compression.
To do this, we will perform a molecular dynamic simulation of a multilayer nanotube system when flat substrates bounding it approach to each other (while reducing the thickness of the packing $\Delta h$).

Let us put our molecular system in a Langevin thermostat at $T=300$~K and start bringing the substrates closer to each other at a constant rate of $v_1=100$~m/s.
To do this, we numerically integrate the Langevin system of equations
\begin{eqnarray}
M\ddot{\bf x}_k=-\frac{\partial E_p}{\partial {\bf x}_k}-\Gamma M\dot{\bf x}_k-\Xi_k,~k=1,\dots,N_{xz},
\label{f10}\\
h_1=v_1t,~~h_2=\Delta h_o-v_1t, \label{f11}
\end{eqnarray}
where ${\bf x}_k$ is $2N$-dimensional vector giving the coordinates of the $k$th nanotube, $E_p$ is potential energy of molecular system (\ref{f8}), $M$ is the mass of carbon atom, $\Gamma=1/t_r$ is the friction coefficient (the relaxation time $t_r=10$~ps),
$$
\Xi_k=\{(\xi_{k,n,1},\xi_{k,n,2})\}_{n=1}^N
$$
is the $2N$-dimensional vector of normally distributed random Langevin forces with the following correlations
$$
\langle\xi_{k_1,n_1,i}(t_1)\xi_{k_2,n_2,j}(t_2)\rangle=2M\Gamma k_BT\delta_{k_1k_2}\delta_{n_1n_2}\delta_{ij}\delta(t_2-t_1)
$$
($k_B$ is Boltzmann constant, $T$ is temperature of the Langevin thermostat).
\begin{figure}[tb]
\begin{center}
\includegraphics[angle=0, width=1.0\linewidth]{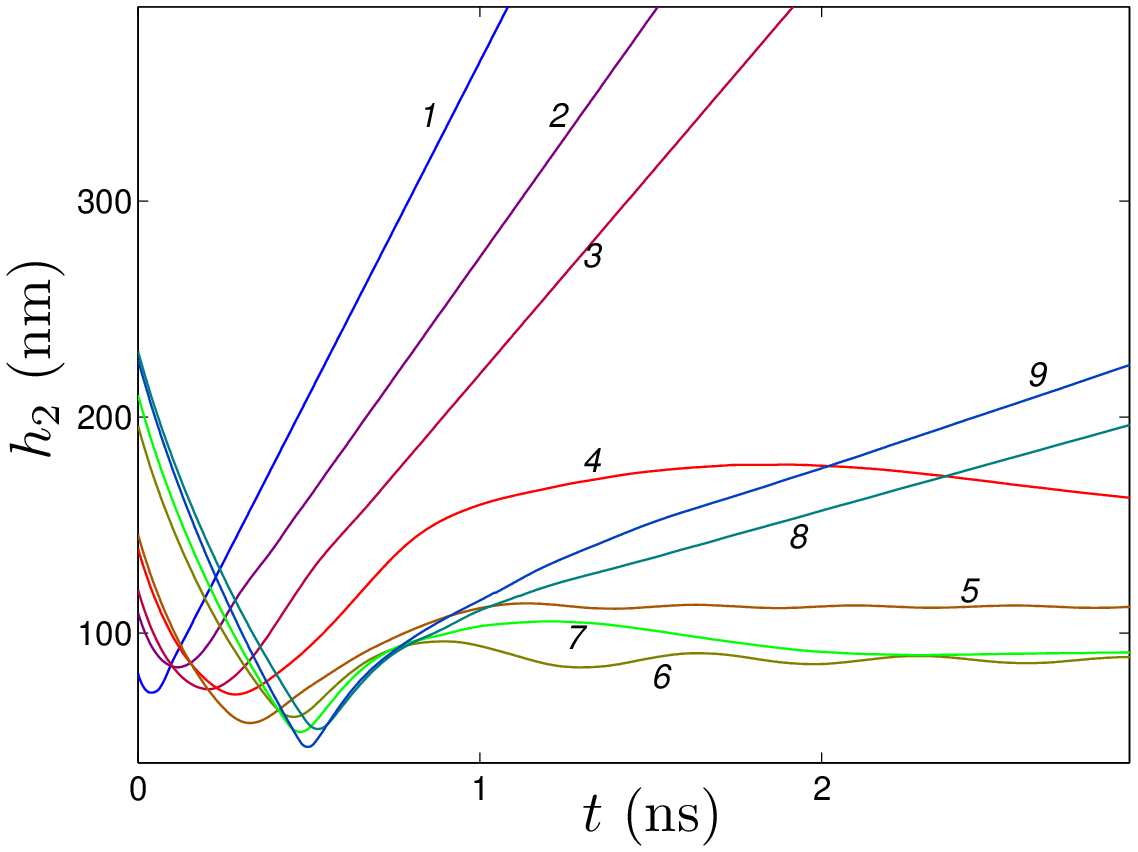}
\end{center}
\caption{\label{fg05}\protect
The dynamics of the upper substrate $h_2(t)$ when the velocity $v_i=600$~m/s is transmitted to it at the initial moment of time ($h_2(0)=\Delta h_o$, $\dot{h}_2(0)=-v_i$) for a multilayer $N_x\times N_z$ CNT $(m,0)$ packing with $m=20$, 30, 35, 40, 45, 55, 65, 70, 80 ( curves 1, 2,...,9).
}
\end{figure}

The system of equations of motion (\ref{f10}) has been integrated numerically using the velocity form of the Verlet difference scheme \cite{Verle1967} with the constant integration step $\Delta t=0.001$~ps.

As an initial condition for the equations of motion (\ref{f10}) we take the stationary state of the layered structure of open nanotubes having a thickness $\Delta h_o$:
$$
\{{\bf x}_k(0)={\bf x}_k^o,~\dot{\bf x}_k(0)={\bf 0}\}_{k=1}^{N_{xz}}.
$$
After reaching the desired packing thickness $\Delta h=\Delta h_o-2v_1t_1$, we fix the position of the substrates, i.e. then we will integrate the system of equations of motion (\ref{f10}) at $h_i(t)\equiv h_i(t_1)$, $i=1,2$.
Further integration allows us to find the average energy value of the layered packing of nanotubes $\bar{E}$ at the selected value of its thickness $\Delta h$:
$$
\bar{E}=\lim_{t\rightarrow\infty}\frac1t\int_{t_1}^{t_1+t}
\left[\sum_{k=1}^{N_{xz}}\frac12(M\dot{\bf x}_k,\dot{\bf x}_k)+E_p(\{ {\bf x}_k\}_{k=1}^{N_{xz}})\right]d\tau .
$$

The change in energy $\bar{E}$ during transverse compression of a multilayer nanotube packing is shown in Fig. \ref{fg03}.
As can be seen from the figure, compression of the CNT$(m,0)$ packing at $m<60$ (diameter $D<4.69$~nm) leads to a monotonous, almost linear increase in energy.
The growth rate decreases monotonously with an increase in $m$ (with an increase in the diameter of the nanotube).
At the index $m=60$, the energy of the packing practically does not change when its thickness decreases from the initial value of $\Delta h_o=190.5$~nm to the value of $\Delta h_1=62$~nm, i.e. in this range, multilayer nanotube packing of under transverse compression behaves like a mechanical structure with zero stiffness.
At $m>60$ (at $D>4.69$~nm), the energy of a multilayer packing decreases during transverse compression and begins to increase only at small values of $\Delta h$, i.e. here the multilayer packing already behaves like a mechanical structure with negative stiffness.
This unusual behavior of multilayer packing is related to the bistability of large-diameter nanotubes, which can be in two stable states: open (cylindrical) and collapsed.
The compression of the packing will occur due to the transfer of part of the nanotubes from the open to the collapsed state \cite{Savin2022pss,Savin2022jetp}. To understand the structural changes of a multilayer packing during compression, we will analyze its possible stationary states.
\begin{figure*}[p]
\begin{center}
\includegraphics[angle=0, width=1.0\linewidth]{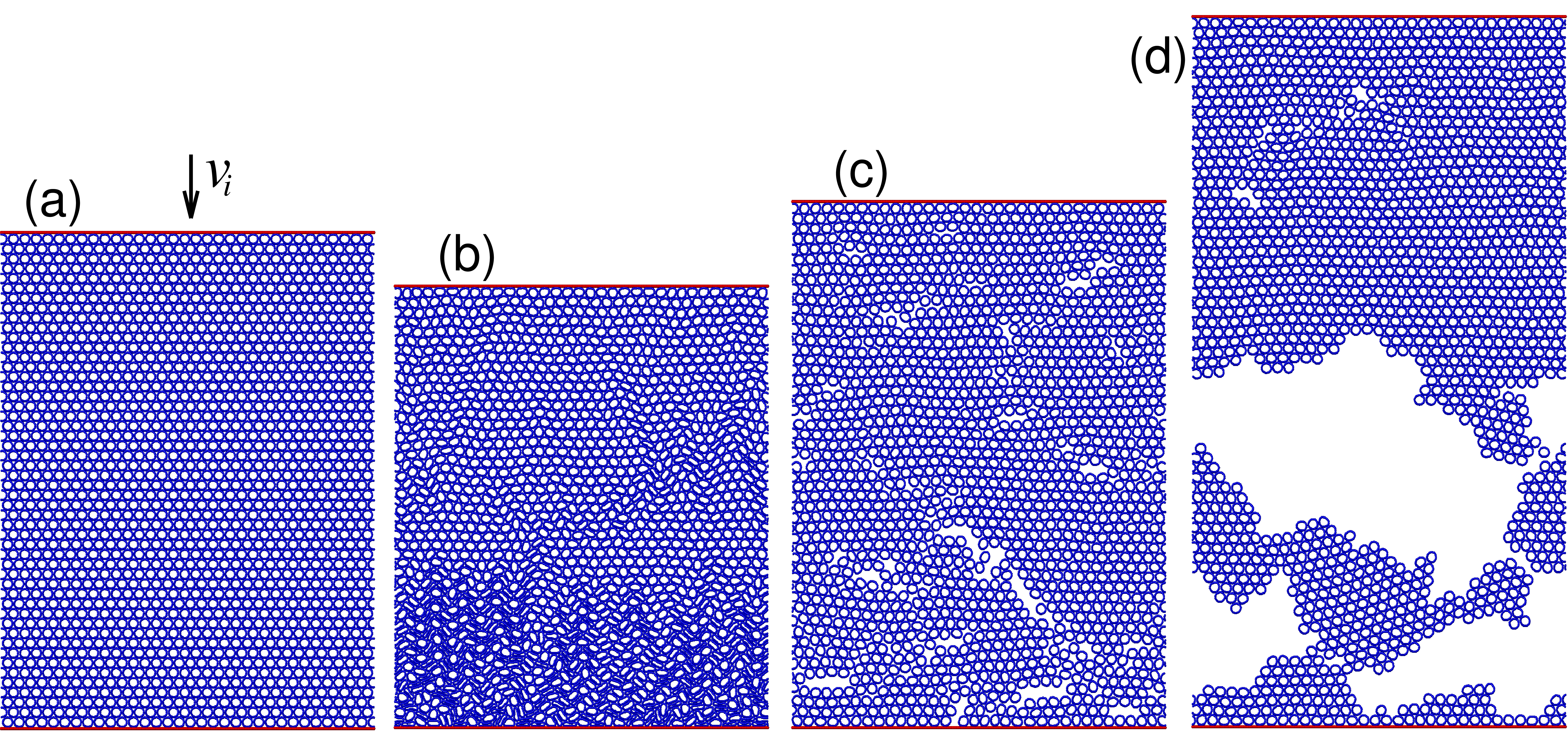}
\end{center}
\caption{\label{fg06}\protect
A change of the structure of the multilayer CNT(20,0) packing  as a result of the transmission of the velocity $v_i=600$~m/s to the upper substrate at the initial moment of time.
Shown are: (a) the initial shape of the packing (time $t=0$);
(b) the shape of the packing at the moment of its greatest compression ($t=t_i=38$~ps);
the shape at (c) $t=100$~ps and (d) $t=200$~ps.
}
\end{figure*}
\begin{figure*}[p]
\begin{center}
\includegraphics[angle=0, width=1.0\linewidth]{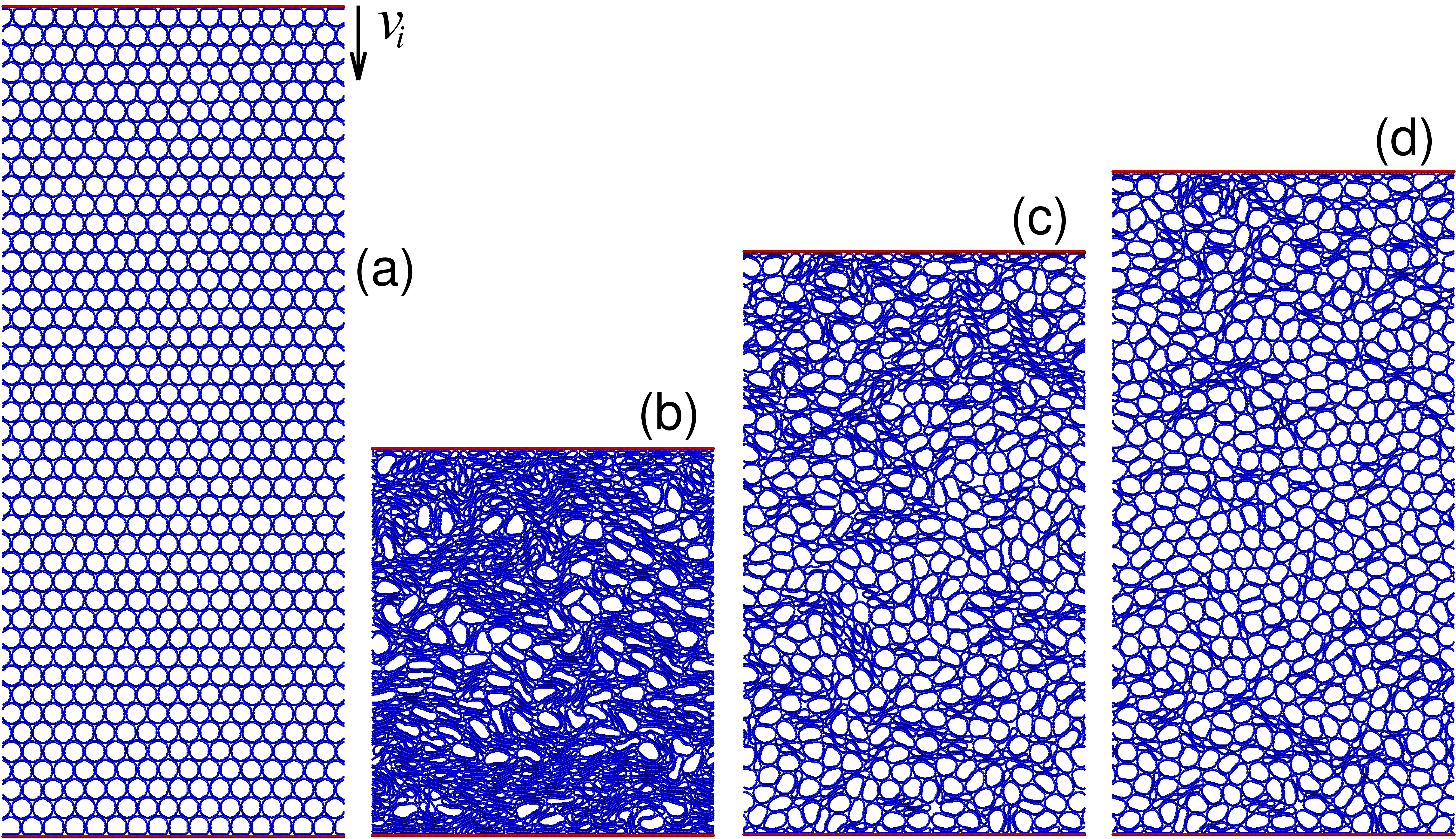}
\end{center}
\caption{\label{fg07}\protect
A change of the structure of the multilayer CNT(45,0) packing as a result of the transmission of the velocity $v_i=600$~m/s to the upper substrate at the initial moment of time.
Shown are: (a) the initial shape of the packing (time $t=0$);
(b) the shape of the packing at the moment of its greatest compression ($t=t_i=232$~ps);
the shape at (c) $t=800$~ps and (d) $t=1600$~ps.
}
\end{figure*}

\section{Stationary states of multilayer CNT packing
\label{sec4}}
An analysis of the possible stationary states of CNTs on a flat substrate \cite{Savin2019ftt} shows that a single-walled nanotube with a chirality index $(m,0)$ has only one stable configuration at $m<32$ and two (open and collapsed) at $m>32$.
Therefore, it should be expected that a multilayer nanotube packing at $m>30$ ($D>2.34$~nm) may have many stable stationary states, which will differ from each other in the proportion of nanotubes in a collapsed state.
As this fraction increases, the thickness of the packing $\Delta h=h_2-h_1$ will decrease monotonously -- see Fig. \ref{fg02}.

To find the stationary states of multilayer CNT packings, we numerically solve the minimum energy problem (\ref{f9}) for a fixed value of the cell period $a_x$.
To obtain all possible stationary packing states, it is convenient to use the configurations obtained by modeling its compression as the starting point in the conjugate gradient method.
Note that when solving the problem (\ref{f9}), the values of the coordinates of the substrates $h_1$ and $h_2$ are not fixed (there is no external compression on the CNT packing).

The solution of the problem (\ref{f9}) has shown that for nanotubes with a chirality index $m<30$ (for nanotubes with $D<2.34$~nm), only one stationary state of multilayer packing is possible -- the state with all open (cylindrical) nanotubes.
Uncompressed stationary states of layered structures with a nonzero fraction of collapsed nanotubes are already possible for nanotubes with $m\ge 30$.
Dependence of the specific energy $E_a=E_p/2mN_{xz}$ (energy per atom) and the proportion of collapsed nanotubes $p=N_c/N_{xz}$ ($N_c$ is the number of collapsed nanotubes in a periodic cell) on the thickness of the stationary state of the packing $\Delta h=h_2-h_1$ shown in Fig. \ref{fg04}.
As can be seen from the figure, in the stationary states of multilayer CNT(30,0) packing, the proportion of collapsed nanotubes can reach a maximum value of $p_c=0.037$, for CNT(35,0) packing -- $p_c=0.811$, for CNT(40,0) -- $p_c=0.989$, and for CNT(45,0) -- $p_c=1$.
For CNTs $(m,0)$ with the index $m\ge 45$, stable states of packings are possible in which all nanotubes are in a collapsed state -- see Fig. \ref{fg02}~(d).

As the number of collapsed nanotubes in a packing increases, its thickness $\Delta h$ decreases monotonously -- see Fig. \ref{fg02} and \ref{fg04}~(b).
Thus, the thickness of the stationary states of the packings can vary significantly from the maximum value of $\Delta h_o$ for the state with all open nanotubes ($p=0$) to the minimum value of $\Delta h_c$ for the state with $p=p_c$.
The values of $\Delta h_o$ and $\Delta h_c$ are shown in the table. \ref{tab1}.

As can be seen from Fig.~\ref{fg04}~(a) for CNTs $(m,0)$ with an index $m<60$ ($D<4.69$~nm), the energy of steady-state of a multilayer packing increases monotonously with an increase in the proportion of collapsed nanotubes $p$ (the collapse of nanotubes does not lead to an energy gain).
For CNTs $(m,0)$ with $m\ge 60$ ($D\ge 4.69$~nm), the packing energy already decreases with an increase in $p$, (the collapse of nanotubes leads to an energy gain).
Here, the multilayer packing behaves like a mechanical system with negative stiffness, its compression leads to a decrease in energy.

Thus, at $35<m<60$ ($2.74<D<4.69$~nm), a multilayer nanotube packing can accumulate energy during compression by collapsing part of the nanotubes and transferring the packing to a stationary state with higher energy.
For nanotubes with a smaller diameter, compression of their multilayer packing occurs elastically without energy accumulation, and for nanotubes with a larger diameter, it occurs with energy release.
Therefore, it can be expected that a multilayer packing of nanotubes with diameters from this range of values will most effectively absorb the energy of an external impact.

\section{Absorption of external impact energy by multilayer CNT packing \label{sec5}}
Let us consider how a multilayer CNT$(m,0)$ packing reacts to an impact on its surface.
To do this, we consider the lower substrate, which has coordinates $h_1=0$, to be stationary, and at the initial moment we inform the upper substrate of the velocity $v_i=600$~m/s (the average value of the bullet velocity).
For the sake of certainty, we consider the mass of the upper substrate to be equal to the mass of all carbon atoms in the periodic cell $M_a=2mN_{xz}M$.
In this case, the dynamics of CNT packing will be set by a system of equations of motion
\begin{eqnarray}
M\ddot{\bf x}_k=-\frac{\partial E_p}{\partial {\bf x}}_k,~k=1,...,N_{xz}, \label{f12}\\
M_a\ddot{h_2}=-\frac{\partial E_p}{\partial h_2},~h_1\equiv 0, \label{f13}
\end{eqnarray}
with the initial condition
$$
\{ {\bf x}_k(0)={\bf x}_k^o,~\dot{\bf x}_k(0)={\bf 0}\}_{k=1}^{N_{xz}},~h_2(0)=\Delta h_o,~\dot{h}_2(0)=-v_i.
$$

The system of equations of motion (\ref{f12}), (\ref{f13}) has been integrated numerically using the velocity form of the Verlet difference scheme with the constant step of integration $\Delta t=0.001$~ps.
The time dependence of the position of the upper substrate $h_2(t)$ is shown in Fig. \ref{fg05}.
Numerical integration of the system of equations of motion has shown that as a result of impact on the upper substrate (sending it a pulse $-M_av_0$), it begins to move downward, compressing the CNT packing, at time $t_i$ maximum compression is achieved, the substrate stops (all impact energy is converted into nanotubes compression energy).
After that, for the packing of CNT$(m,0)$ with $m\le 35$, an elastic restoration of the shape of the nanotubes occurs, leading to a monotonous reverse movement of the substrate.
As a result, the substrate detaches from the packing, taking part of the nanotubes with it -- see Fig. \ref{fg05} (curves 1, 2, 3) and \ref{fg06} ($h_2(t)\nearrow\infty$ for $t\nearrow\infty$).
Here, an impact to the packing leads to its destruction.

For the CNT$(m,0)$ packing with $45\le m\le 65$, an impact to the upper substrate does not destroy the packing, as a result of the collapse of a part of the nanotubes, it passes into a higher-energy stationary state -- see Fig. \ref{fg05} (curves 4-7) and \ref{fg07}.
This is where the packing completely absorbs the impact energy.

For the CNT$(m,0)$ packings with $m\ge 70$, the impact already leads to their destruction.
As a result of the reverse movement of the substrate, it is completely detached from the packing with part of the nanotubes -- see Fig. \ref{fg05} (curves 8, 9).
\begin{figure}[tb]
\begin{center}
\includegraphics[angle=0, width=1.0\linewidth]{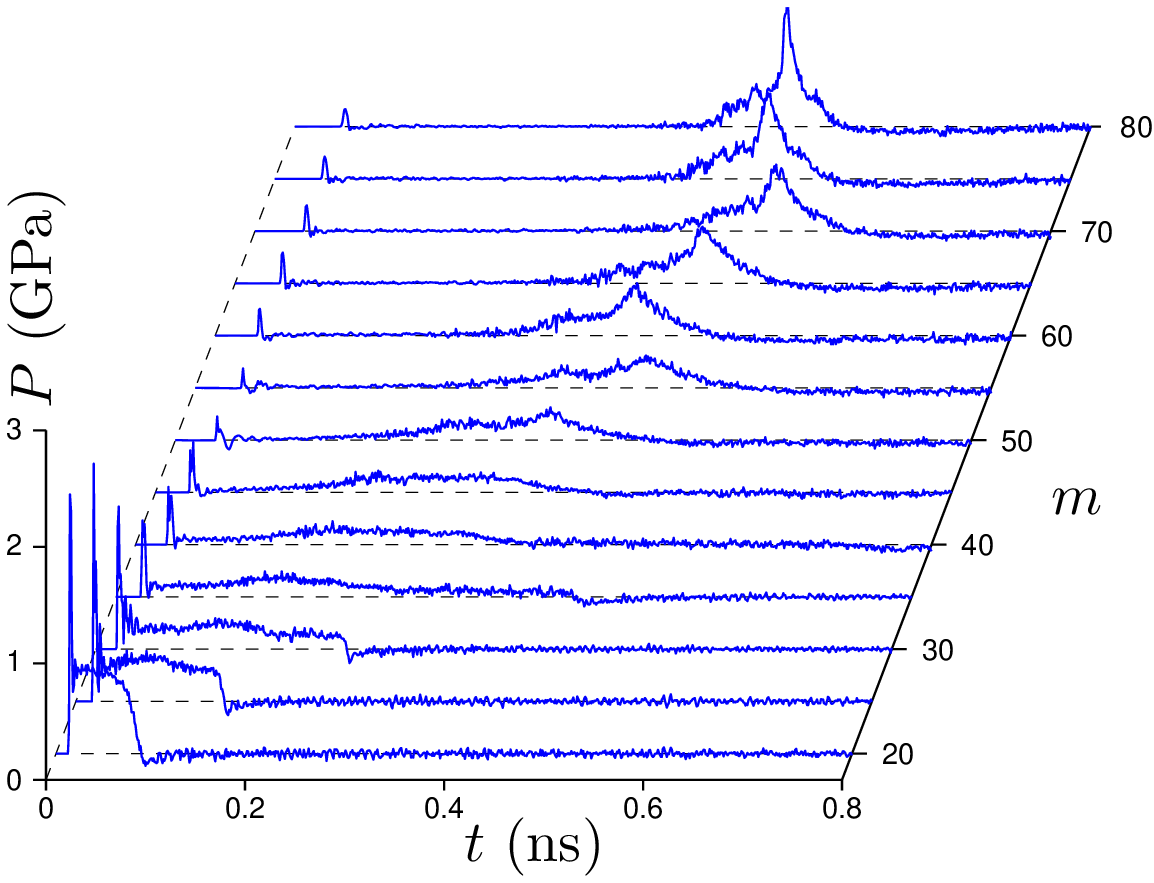}
\end{center}
\caption{\label{fg08}\protect
Time dependence $t$ of the pressure $P$ on the lower substrate of a multilayer CNT$(m,0)$ packing resulting from the velocity $v_i=600$~m/s being transmitted to the upper substrate at the initial moment of time.
}
\end{figure}

The main task of the shock-absorbing layer is to partially absorb the impact energy and distribute the initially narrow dynamic pulse over the longest possible time interval, so that a weakened pulse spread over a long time interval reaches the lower (stationary) substrate.
Let's consider how the multilayer CNT$(m,0)$ packing copes with this task.
To do this, we will analyze the shape of the pulse reaching the lower substrate.
The lower stationary substrate perceives the instantaneous impact on the upper substrate as additional pressure arising
$$
P(t)=\frac{1}{S}\frac{\partial E_p}{\partial h_1},
$$
where $S=a_x\Delta_y=3a_xr_c/2$ is the area of the substrate.

The dependence of pressure $P$ on time $t$ is shown in Fig. \ref{fg08}.
As can be seen from the figure, the additional pressure resulting from the impact is manifested in the finite time interval $t_1\le t\le t_2$.
At time $t_1$, a shock wave arrives to the lower substrate (elastic deformations of nanotubes).
The motion of such a wave under the shock load of a bundle of single-walled carbon nanotubes in the transverse direction was modeled in \cite{Galiakhmetova2022}.
The pressure peak associated with this wave is most pronounced for small-diameter nanotube packings [for CNTs $(m,0)$ with $m\le 30$, diameter $D<2.35$~nm].
An increase in the index $m$ (an increase in the diameter of the nanotubes $D$) leads to a monotonous decrease in this peak.
After this peak at $m\le 35$, a longer pressure effect occurs almost immediately due to the arrival of a plastic deformation zone.
Here the pressure amplitude is much lower, but its duration is longer.
For CNT packings with a larger diameter, the pressure from the arrival of plastic deformations appears much later than the arrival of the elastic wave.
For these packings, the main contribution to the pressure on the fixed substrate is associated with the arrival of plastic deformations.
The lowest maximum pressure from these deformations is observed for nanotubes with index $m=40$ and 45.

\begin{figure}[tb]
\begin{center}
\includegraphics[angle=0, width=1.0\linewidth]{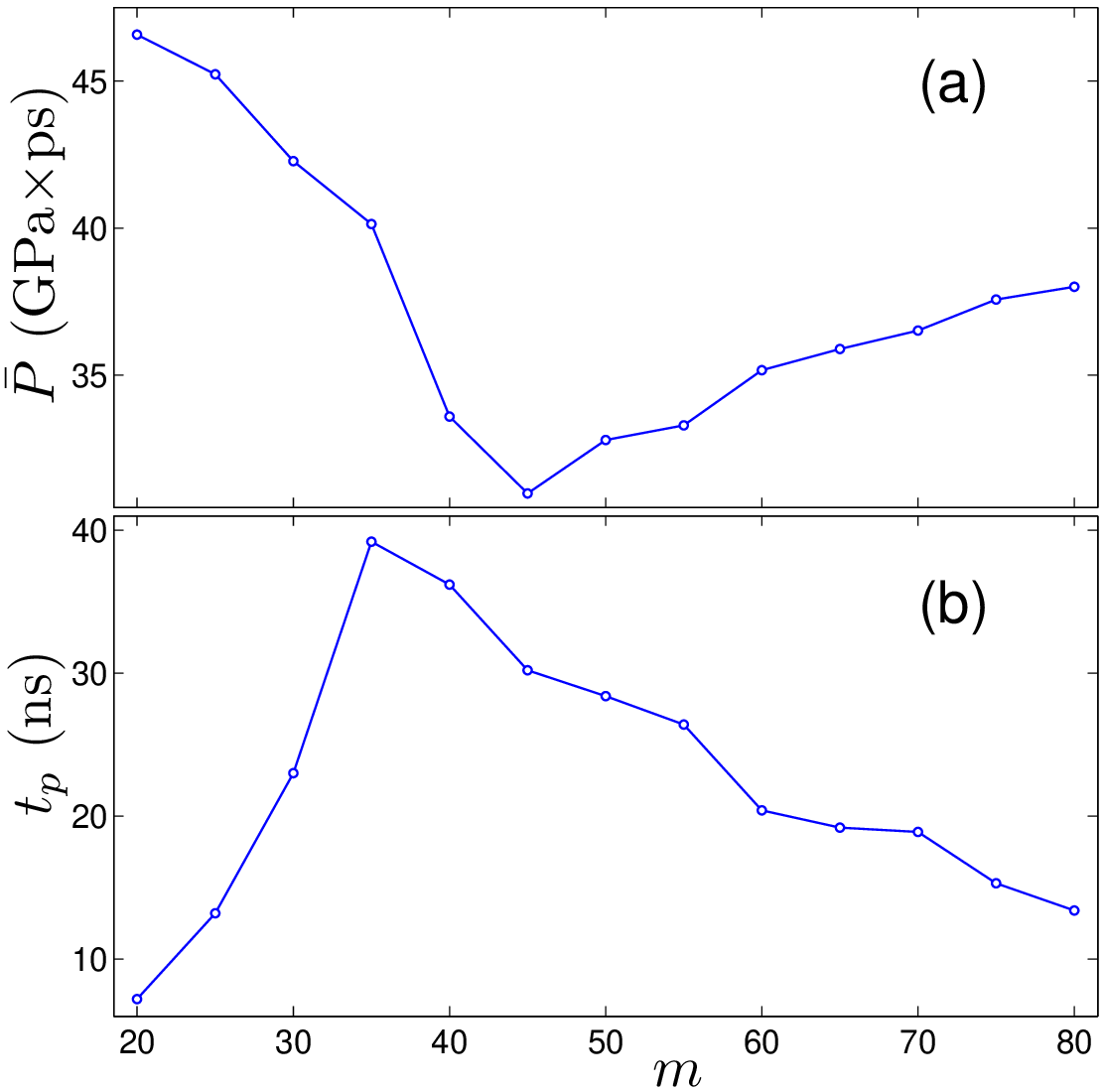}
\end{center}
\caption{\label{fg09}\protect
The dependence (a) of the pressure impulse $\bar{P}$ and (b) the time of its action $t_p$ on a fixed substrate on the value of the chirality index of nanotubes $m$ of a multilayer packing.
}
\end{figure}

To analyze the magnitude of the impact on a stationary substrate, it is convenient to introduce an additional pressure impulse $\bar{P}=\int_0^{t_2}P(t)dt$ and an impact time $t_p=\int_0^{t_2}\theta(P(t))dt$, where the function $\theta(P)=1$ for $P>0.03$ and $\theta(P)=0$ for $P<0.03$~GPa.
The dependence of $\bar{P}$ and $t_p$ on the chirality index value $m$ of nanotube multilayer packings  is shown in Fig. \ref{fg09}.
As can be seen from the figure, the pressure impulse $\bar{P}$ on a stationary substrate will be the smallest upon impact on multilayer CNT(45,0) packing, and the impact time $t_p$ will be maximum at $35\le m\le 50$, i.e. for an array of nanotubes with diameter $2.7<D<3.9$~nm.

Thus, the simulation shows that multilayer packings of single-walled nanotubes with a diameter of $D\in [2.7,3.9]$~nm will best serve as a shock-absorbing layer.
It is for these nanotubes that the absorption of impact energy will occur most strongly due to the transfer of the packing to a higher energy steady state, and the impact pulse will reach the lower substrate in the most weakened and time-distributed manner.

Note that simulations of a collision between a suspended graphene sheet and a nanotube moving at a speed of $v=3000$~m/s show \cite{Yang2016} that the collision leads to the appearance of wave fronts of longitudinal and transverse vibrations in the sheet, which carry away up to 90\% of the impact energy.
A graphene sheet allows the energy of a narrow impact to be redistributed over an area.
It should be expected that the structure consisting of alternating layers of single-walled CNTs with a diameter of $2.7<D<3.9$~nm and graphene sheets will serve even better as a shock-absorbing layer.
Here, the nanotube layers will absorb the impact energy and spread the passing pulse over time, while the graphene sheets will ensure its redistribution over the area.

\section{Conclusion \label{sec6}}

The simulation has shown that the multilayer packings (arrays) of parallel single-walled carbon nanotubes can act as effective shock absorbers.
The depreciation effect will be most pronounced for packings with nanotube diameter of 2.7 -- 3.9~nm.
For the packing of such nanotubes, a portion of the impact energy is absorbed.
The packing passes into a higher-energy steady state due to the collapse of some of the nanotubes.
The impact impulse will reach the other edge of the packing most weakened and distributed over time.
For nanotubes of smaller diameter, the compression of their packing occurs elastically without energy accumulation, and for nanotubes of larger diameter, the compression already occurs with the release of energy.

The efficiency of energy absorption is due to the multistability of the packings of medium-diameter nanotubes -- the presence of higher-energy stationary states.
These states differ from each other by the proportion of nanotubes in the collapsed state.
Such packages behave like nanoscale honeycomb structures (metamaterials) with negative stiffness \cite{Darwish2024,Tan2025}.
Note that if the layers of nanotubes alternate with graphene sheets separating them, then such a layered structure will be an even more effective shock absorber.
Here, the layers of nanotubes will absorb the impact energy and spread the passing pulse over time, and the graphene sheets will redistribute it over the area.
\\ \\
{\bf Acknowledgements}\\

The research was funded by the Russian Science Foundation (RSF) (project No. 25-73-20038).
\\ \\
%

\end{document}